\documentclass[aip,rsi,preprint,graphicx,superscriptaddress]{revtex4-1} 

\usepackage{amsmath}
\usepackage{amssymb, bbm, graphicx,color}

\begin{document}
\title{ Generalized lock-in amplifier for precision measurement of high frequency signals}

\author{Siyuan Fu}
\affiliation{Department of Electrical and Information Technology, Lund University, Ole R\"omers v\"ag 3, 22363, Lund, Sweden}
\author{Atsunori Sakurai}
\affiliation{Chemical Physics, Lund University, Getingev\"agen 60, 22241, Lund, Sweden}
\author{Liang Liu}
\affiliation{Department of Electrical and Information Technology, Lund University, Ole R\"omers v\"ag 3, 22363, Lund, Sweden}
\author{Fredrik Edman}
\affiliation{Department of Electrical and Information Technology, Lund University, Ole R\"omers v\"ag 3, 22363, Lund, Sweden}
\author{T\~onu Pullerits}
\affiliation{Chemical Physics, Lund University, Getingev\"agen 60, 22241, Lund, Sweden}
\author{Viktor \"Owall}
\affiliation{Department of Electrical and Information Technology, Lund University, Ole R\"omers v\"ag 3, 22363, Lund, Sweden}
\author{Khadga Jung Karki}
\email{Khadga.Karki@chemphys.lu.se}
\affiliation{Chemical Physics, Lund University, Getingev\"agen 60, 22241, Lund, Sweden}





\begin{abstract}
We herein formulate the concept of a generalized lock-in amplifier for the precision measurement of high frequency signals based on digital cavities. Accurate measurement of signals higher than 200 MHz using the generalized lock-in is demonstrated. The technique is compared with a traditional lock-in  and its advantages and limitations are discussed. We also briefly point out how the generalized lock-in can be used for precision measurement of giga-hertz signals by using parallel processing of the digitized signals. 
\end{abstract}


\maketitle

\section{Introduction}
The concept of the digital cavity~\cite{KARKI2013A} that has been recently introduced mimics the functionality of an analog cavity. Cavities, in general, use the interference effect of the electromagnetic (EM) fields to generate comb-filters. The number of cycles 
that interfere determine the line-width of the cavity.~\cite{VAUGHAN_1977} The ratio of the line-width to the frequency of the signal is often termed the ``Q-factor'' of the cavity. The losses in a cavity due to absorption and scattering limit the number of free oscillations of the EM field in the cavity thereby limiting the number of cycles that interfere, and consequently the Q-factor of the cavity. The highest Q-factor achievable in an analog design is about 10$^6$.~\cite{COLLIN_2001} 

The power loss in the cavity relates to the loss in the information. Alternatively, one can design a detection system such that the complete information of the EM field gets stored before it is lost using an analog to digital converter (ADC). This is possible for signals up-to a few tens of giga-hertz (microwave frequencies) using ADCs on the market. Once the signal has been digitized one can use a digital cavity to mimic the interference effect of an analog cavity. In this case, as no information is lost due to the power loss, a digital cavity with extremely narrow line-width can be designed. The algorithms of such filters have very low computational foot-print and can be parallelised, which opens up the possibility of implementing real-time phase sensitive detection (PSD) schemes for microwave signals with extremely high Q-factors. We discuss one of such possibilities in this article, viz. generalization of the lock-in amplifier (LIA).  

\section{Theory}
The response of a digital cavity is described by the following equation:~\cite{KARKI2013A}
 
\begin{equation}\label{EQ1}
y(j\cdot \Delta t;n):=\sum_{k=0}^{N_c} x(j\cdot \Delta t+k\cdot T); j=1,2,...n,
\end{equation}
where $y$ is the response of the cavity to the signal $x$ and $\Delta t$ is the digitization interval. According to the above equation, the digitized signal $x$ is sectioned to $N_c+1$ waveforms of $n$ elements and those waveforms are then added to generate the response.  This simple algorithm captures the interference effect of the sinusoidal signals, i.e. only the frequencies whose periods match the length of the cavity, $n\times \Delta t$, interfere constructively while the other frequencies interfere destructively. For large $N_c$, the algorithm generates a comb-pass filter with line-width given by 
\begin{equation}\label{EQ2}
2\xi_{f_0} \leq \frac{2 \sqrt{2}f_0 }{N_c+1},
\end{equation}
where $2\xi_{f_0}$ is the full-width at half maximum of the response of the cavity and 
\begin{equation}\label{EQ3}
f_0=1/(n\cdot \Delta t)
\end{equation}
 is the fundamental resonance frequency of the cavity.~\cite{KARKI2013A} According to Equ.\eqref{EQ3} the fundamental resonance frequency can be tuned by either changing $n$ or $\Delta t$. For example, if a signal is digitized at 100 MS/s (mega samples per second) and $n$ is chosen to be 100 then $f_0=1$ MHz, and if $n$ is chosen to be 90 then $f_0=1.1111...$ MHz. In this case $\Delta t = 0.01 \mu$s. For high resolution signal analysis changing $n$ provides a coarse tuning while changing $\Delta t$  provides a fine tuning of the digital cavity. $\Delta t$ is defined by the clocking of the ADCs. It can be changed continuously over a certain range by using a  high precision voltage controlled oscillator. This design of the digital cavity is expected to achieve extremely high Q-factors  $> 10^9$ with reasonable memory requirements.~\cite{KARKI2013A} A drawback of such a design is that  normal digitizing cards that usually have a fixed clock cannot be used to build a finely tunable digital cavity.

 However, in applications for which a Q-factor of less than $10^8$ suffices one can use digital tuning.  Digital tuning refers to the concept of shifting the frequency of the signal to the resonant frequency of the digital cavity by multiplying it with an appropriate sine or cosine function. Similar technique is also used in the LIA where it forms a part of the PSD scheme.~\cite{FREITAS_1979} A common digital LIA uses two 
references, one in-phase and other quadrature:~\cite{MARQUES_2004}
\begin{equation}\label{EQ4}
V_{\textrm{RP}} = \cos(\omega_{\textrm{ref}}\cdot t), \,\,\, \textrm{in-phase}
\end{equation}
\begin{equation}\label{EQ5}
V_{\textrm{}RQ} = \sin(\omega_{\textrm{ref}} \cdot t), \,\,\, \textrm{quadrature}
\end{equation}
to demodulate the signal. The multiplication of the signal $V_{\textrm{sig}} = A \sin(\omega_{\textrm{sig}}\cdot t + \phi)$ by the reference signal(s) produces side bands at the sum and difference frequencies, e.g. multiplication  by the in-phase reference gives

\begin{equation}
V_{\textrm{SP}} = \frac{A}{2} \left[ \sin ((\omega_{\textrm{sig}}-\omega_{\textrm{ref}})t+\phi)+ \sin((\omega_{\textrm{sig}}+\omega_{\textrm{ref}})t+\phi)\right],
\end{equation}
 where $V_{\textrm{SP}}$ is the in-phase signal after multiplication with the reference. 
When $\omega_{\textrm{sig}} = \omega_{\textrm{ref}}$, the resulting signal contains a DC component, $V_{\textrm{SP-DC}} = A \sin(\phi)/2$, and an AC component , $V_{\textrm{SP-AC}} = A \sin((\omega_{\textrm{sig}}+\omega_{\textrm{ref}})t+\phi)/2$. A sharp cut-off low-pass filters are used to select the DC components of the in-phase and the quadrature signals from which the phase and the amplitude of the signal are calculated. The reference signal and the low-pass filter play critical role in the LIA.~\cite{MARQUES_2004,WU_1989, JAQUIER_1994, RUSSO_1995,  BONETTO_2005} The algorithms for generating the in-phase and the quadrature references usually involve computationally costly digital phase-locked-loops (PLL), and the same is true for the design of digital low-pass filters with high Q-factors.~\cite{MARQUES_2004,BONETTO_2005} This limits the use of common digital LIA for precision measurements of signals to frequencies lower than few hundred MHz. New techniques that avoid using reference signals and use computationally efficient narrow line-width digital filters with parallelisable algorithms for computation are necessary to precisely measure the high frequency signals that exceed the clock frequency of the computer or the digital signal processing  (DSP) unit. Digital cavities fulfill these requirements.   

When using a digital cavity for high precision signal analysis one uses the line filter at the fundamental resonant frequency $\omega_0=2\pi f_0$. The signal is multiplied by a cosine (or sine)  function, $\cos(\omega_{\textrm{shift}} t)$, to get the input function, $V_{\textrm{Input}}$, to the cavity:
\begin{equation}\label{EQ6}
V_{\textrm{Input}} =  \frac{A}{2} \left[ \sin ((\omega_{\textrm{sig}}-\omega_{\textrm{shift}})t+\phi)+ \sin((\omega_{\textrm{sig}}+\omega_{\textrm{shift}})t+\phi)\right].
\end{equation} 
$\omega_{\textrm{shift}}$ is chosen such that one of the frequency components of $V_{\textrm{Input}}$ is resonant with the cavity. In practice, it is better to chose $\omega_{\textrm{shift}}$ such that $\omega_{\textrm{sig}}+\omega_{\textrm{shift}} = \omega_0$. In this case the frequency component at $\omega_{\textrm{sig}}-\omega_{\textrm{shift}}$  is always non-resonant with the cavity for non-zero $\omega_{\textrm{shift}}$. Care must be taken when chosing $\omega_{\textrm{sig}}-\omega_{\textrm{shift}}=\omega_0$ as in this case $\omega_{\textrm{sig}}+\omega_{\textrm{shift}}$ can also be equal to the multiples of $\omega_0$ and be resonant with the cavity.

 The output of the digital cavity is the amplified waveform, $(A/2) (N_c+1) \sin((\omega_{\textrm{sig}}+\omega_{\textrm{shift}})t+\phi)$, from which the amplitude $A$ and phase $\phi$ of the signal can be extracted. The waveform averaging generates the narrow line-pass filter, and it also reduces the white noise in the signal by a factor of $\sqrt{N_c+1}$.~\cite{KARKI2013A}  

\section{Test of a generalized lock-in  }
\subsection{Precision measurements of high-frequency signals}
In this test, a signal at 245 MHz was generated using the signal generator HP-8656B from Hewlett-Packard. An 8 bit digitizer ATS9840 from Alazartech was used to digitize the signal at the rate of 1 GS/s (giga sample per second). A digital cavity was set with $n=4$, for which the $f_0=250 MHz$. The digitized signal was frequency up-shifted by multiplying with  sine functions with $f_{\textrm{shift}}$ around 5 MHz and filtered with the digital cavity to get the spectral information in the vicinity of $245$ MHz. 

\begin{figure}[h]
\includegraphics[width=5in]{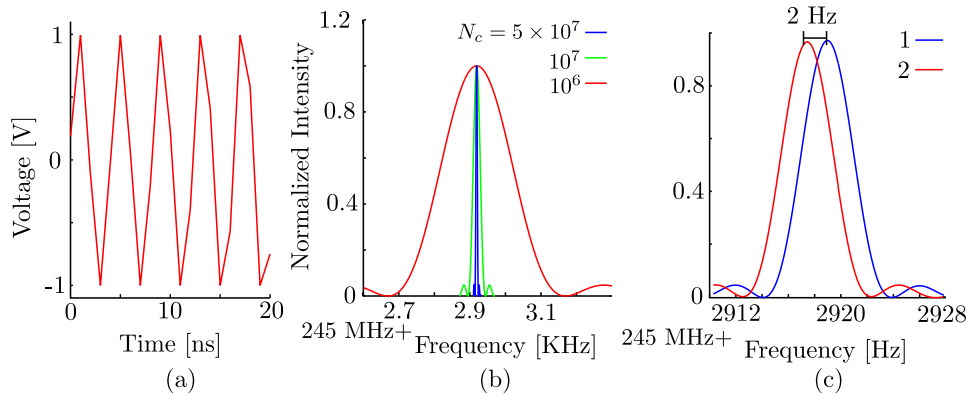}
\caption{Precision measurement of the high-frequency signals using generalized lock-in. (a) is the raw signal, (b) is the response of the digital cavity with varrying cavity folds, $N_c$ to the upshifted signal and (c) is the comparison of the cavity response with $N_c=5\times 10^7$ to the signals acquired at 0.5 s time delay. }
\label{FIG1}
\end{figure}

Fig. \ref{FIG1}a shows the signal acquired by the digitizer. Fig. \ref{FIG1}b shows the spectrum of the signal around 245 MHz. The scanning of the spectrum is done by up-shifting the signal to the resonant frequency of the digital cavity at 250 MHz. Though the signal generator is set to 245 MHz Fig. \ref{FIG1}b shows that the signal measured by the digitizer is at 245.0029 MHz. This discrepancy is due to the slightly different clock speeds of the generator and the digitizer. The response of the digital cavity with $N_c$ up $10^7$, shown by red and green curves in Fig. \ref{FIG1}b, is close to the response of the cavity to an ideal monochromatic signal.~\cite{KARKI2013A} However, when $N_c$ is increased to $5\times 10^7$, one observes wobbling of the maxima of the cavity response by few Hz. Fig. \ref{FIG1}c shows a 2 Hz shift of the maxima of the cavity response in the signals that are acquired half a seconds apart. The wobbling of the frequency of the signal is due to the fluctuations in the signal generator caused by temperature drifts, vibrations, etc.

\subsection{Signal recovery from large background noise}
\begin{figure}
\includegraphics[width=3.25in]{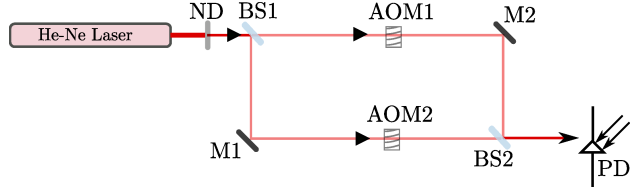}
\caption{Schematic of the setup for recording the noisy data using photo-diode (PD). The laser beam from the He-Ne laser is split into two beams using a beam splitter BS1. The phase of each of the beams is modulated using acousto-optic modulators (AOMs) and combined again using another beam splitter BS2. The intensity of the combined beam is modulated at the difference frequency of the AOMs.  }
\label{FIG2}
\end{figure}

In this test we recorded the signal from a photodiode using a 14 bit digitizer ATS9440 from Alazartech.
The experimental setup is shown in Fig. \ref{FIG2}. A continuous-wave He-Ne laser (25LHP121-230) from CVI Melles Griot was used as the light source. In the measurements, the light from the source is attenuated using a neutral density filter (ND, OD$\approx$2) and split into two beams using a beam splitter (BS1). The phase of each of the beam is modulated by using acousto-optic modulators (AOM1,2) from Gooch and Housego (R35055-1-.8). The two beams are then recombined using another beam splitter (BS2). The recombined beam is then detected by an amplified photo-diode (PD, PDA36A Thorlabs). The phase modulation by AOMs causes the intensity modulation of the combined beam at the difference frequency of the phase modulation. The phase modulation frequency of the AOMs is varied in-between 40-70 MHz to get the difference frequency of up to 30 MHz. 

\begin{figure}
\includegraphics[width=5.0in]{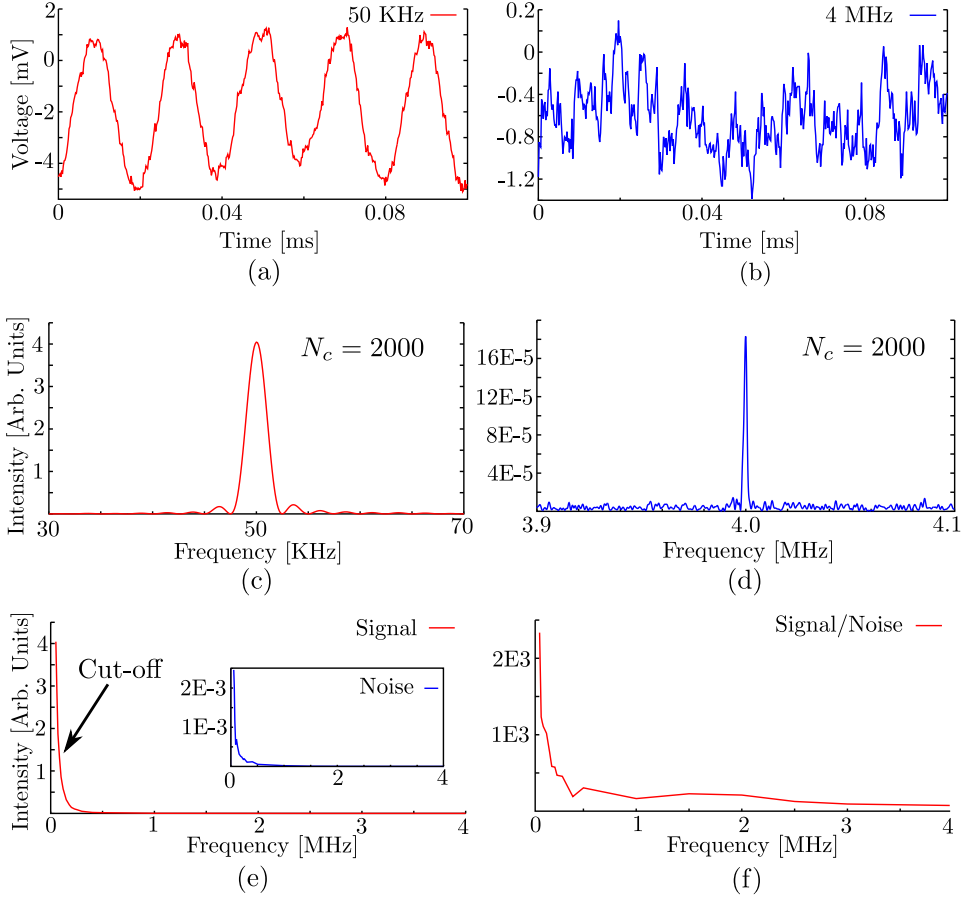}
\caption{Photo-diode signal for modulated light. (a) shows the signal at 50 KHz, (b) shows the noisy signal at 4 MHz, (c) shows the digital cavity response to the 50 KHz signal, (d) shows the similar response from the 4 MHz signal,  (e) shows the photo-diode signal vs. the modulation frequency of the light and (f) shows the signal to noise ratio for the different modulation frequencies. The maximum of the cavity response as shown in (c) and (d) for the measurements of various signals between 50 KHz to 4 MHz is taken as the intensity in (e).}
\label{FIG3}
\end{figure}

Fig. \ref{FIG3}a-b show the photo-diode signals for the light modulated at 50 KHz and 4 MHz, respectively. The signals are digitized at the rate of 50 MS/s. The response of a photo-diode depends on the modulation frequency of the light. The amplifier used in the photo-diode typically has a cut-off frequency beyond which the amplification of the photo-diode response falls down rapidly. Since 4 MHz lies beyond the cut-off frequency of about 100 KHz (shown in Fig. \ref{FIG3}e) the photo-diode signal is very noisy at these frequencies. We use a digital cavity set to 5 MHz ($n=10$ and $N_c=2000$), and up-shift the signals recorded by the photo-diode to the resonant frequency of the digital cavity by multiplying it with sine functions having appropriate frequencies, $\omega_{\textrm{shift}}$ (e.g. for 50 KHz signal $\omega_{\textrm{shift}}=4.05 MHz$ and for 4 MHz $\omega_{\textrm{shift}}=1 MHz$. Fig. \ref{FIG3}c-d show the response of the digital cavity to the signal around 50 KHz and 4 MHz, respectively. The scanning for the signals around 50 KHz and 4 MHz is done by changing $\omega_{\textrm{shift}}$. The signal at 50 KHz, Fig. \ref{FIG3}a has little noise, consequently the response of the digital cavity is close to the ideal noise free signal~\cite{KARKI2013A}, while the response of the cavity to 4 MHz signal shows some contribution due to noise around the central frequency. 

The amplitudes of the response of the digital cavity for photo-diode signals at different modulation frequencies is shown in Fig. \ref{FIG3}e. The inset in Fig. \ref{FIG3}e shows the noise of the photo-diode at the different frequencies.  The amplitudes of the signals at higher frequencies is negligible compared to the signals at the lower frequencies (Fig. \ref{FIG3}e), nevertheless  the signals at higher frequencies are easily discernible from the noise (Fig. \ref{FIG3}d). As shown in Fig. \ref{FIG3}f the noise at the lower frequencies is about 3 orders of magnitude smaller than the signal after the application of digital cavity, while it is about 80 times smaller in the case of the higher frequencies.   

\begin{figure}
\includegraphics[width=5in]{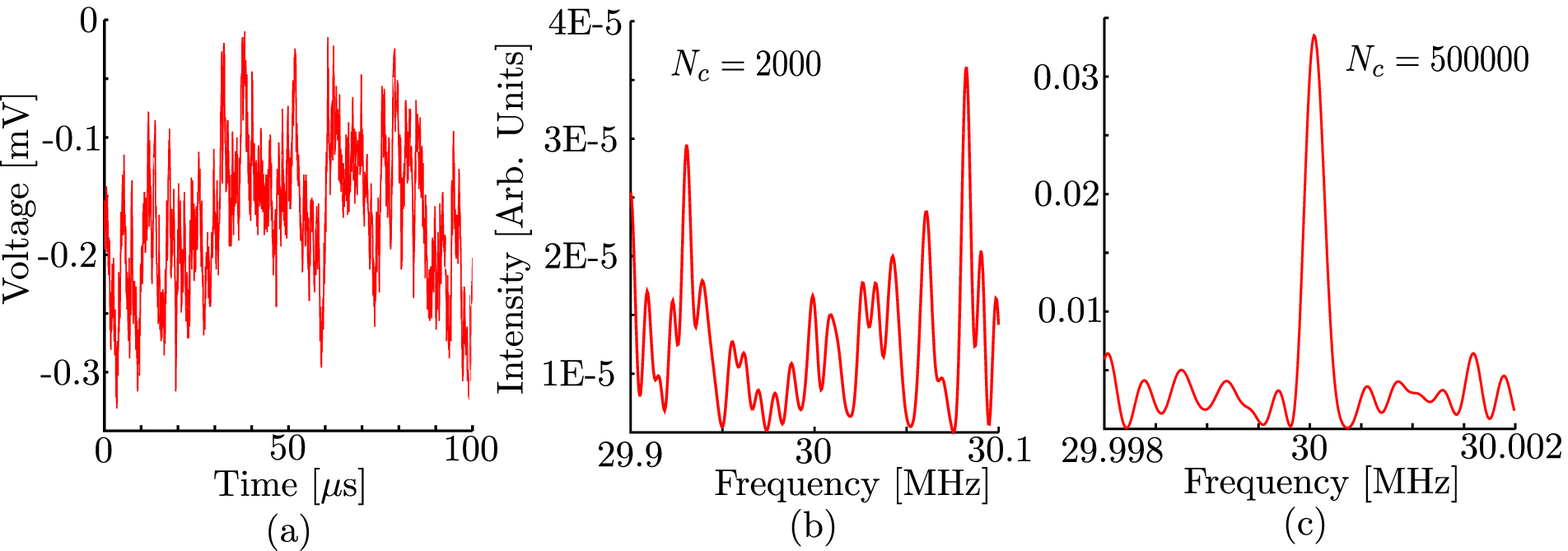}
\caption{Signal retrieval at very high frequencies. (a) shows digitized signal, (b) shows the amplitude of the signal around 30 MHz after applying a digital cavity with $N_c=2000$ and (c) shows the amplitude of the signal around 30 MHz after applying a digital cavity with $N_c=500000$. }
\label{FIG4}
\end{figure}

It is possible to use the generalized lock-in to retrieve signals even at higher frequencies. Fig. \ref{FIG4} shows an example of the signal retrieved at 30 MHz. Fig. \ref{FIG4}a shows a part of the actual signal digitized by the digitizer at the rate of 100 MS/s. The data is predominantly electronic noise. The signal contribution to the digitized data at such high frequency is so low that it cannot be isolated from the noise even after applying a digital cavity set to 25 MHz and $N_c=2000$ as shown in Fig. \ref{FIG3}. The signal at 30 MHz is visible only when we use a digital cavity with very high $N_c$ (Fig. \ref{FIG4}c). Though we have tested signal recovery from extremely noisy data for signals up to 30 MHz, generalized lock-in can be used for even higher frequencies provided that one uses suitable digitizer and appropriate values of $N_c$. 30 MHz is the highest modulation frequency we can generate in our setup.  

\subsection{Advantages and limitations}
The development of a generalized LIA is motivated by the need for precision analysis of high frequency signals. The generalized LIA simplifies the algorithms for the line-filters and also eliminates computationally demanding PLLs. As the algorithms are parallelisable, this technique can be used to analyze frequencies beyond the clock cycles of the DSP processors. With the most advanced commercially available digitizers and on-board processors, the generalized LIA could be developed to analyze the signals upto 60-70 GHz.

A generalized LIA does not use an external reference signal, which is an advantage as well as a limitation. The noise in the phase and the amplitude of the reference in an LIA add to the noise in the output.~\cite{JAQUIER_1994,MARQUES_2004} This source of noise is eliminated in the generalized LIA. As the generalized LIA is not locked to a reference, it is freely tunable. This is also an advantage because one can apply multiple cavities simultaneouly to the signal to extract different Fourier components.~\cite{KARKI2013A} Moreover, for the applications like the software defined radio or the detection of faint radio/microwave frequencies the external reference is not available so normal LIA cannot be used.

In the case when the generalized LIA is used to measure only one Fourier component in the signal, the phase of the Fourier component is the phase when the digitization commences, so it is a random number. This is the limitation of not using the external reference.  One of the ways to get a meaningful phase information uses a two channel digitizer,~\cite{WU_1989} one of which digitizes the signal and the other digitizes the reference. The generalized LIA is used to extract the phase and amplitude of the both. The phase difference between the signal and the reference then is independent of the starting time of the digitization. This implementation of the generalized LIA is suited to analyze the phase and amplitude of the noisy signals in optical non-linear spectroscopy.~\cite{MARCUS2006, MARCUS2007,KARKI2012A, KARKI2012B}

The generalized LIA uses a comb-pass filter. This means it selects not only a single Fourier component of the signal at the frequency $\omega_{\textrm{sig}}$ but all the other components at $\omega_{\textrm{sig}}+\omega_0, \omega_{\textrm{sig}}+2\omega_0, ... $ etc. This is an advantage in certain situations when the harmonics of the signal needs to be analyzed but is a disadvantage when a single Fourier component needs to be filtered. The disadvantage, however, can be eliminated in different ways. One of the ways is to use the high-frequency cut-off of the digitizers themselves for signal pre-conditioning such that higher frequencies do not reach the ADC. For example, when a digitizer with bandwidth 450 MHz is used for the generalized LIA, one can set $\omega_0 = 250$ MHz. 

\begin{figure}
\includegraphics[width=5in]{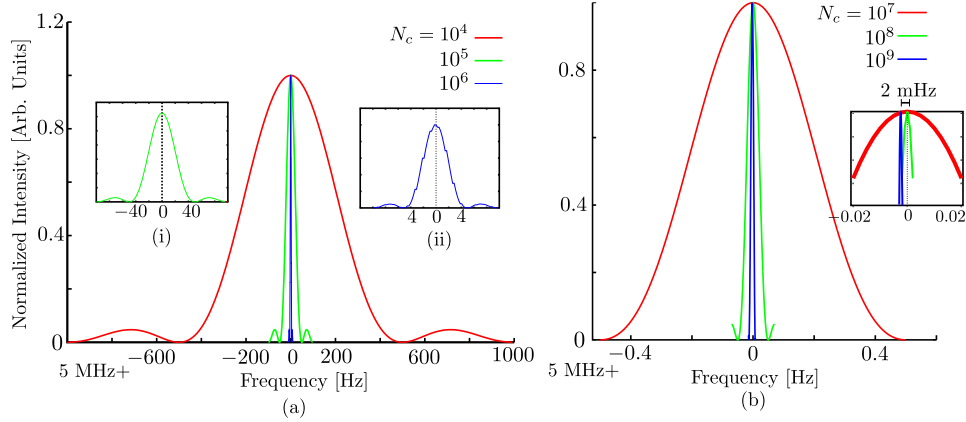}
\caption{Cavity response to the pure sine signals generated by the computer around 5 MHz. 32 bit floating number system is used in (a) and 128 bit number system is used in (b).  }
\label{FIG5}
\end{figure}

The generalized LIA uses the computer system's built-in sine and cosine functions, which have finite numerical precision, and thus makes the accuracy of the algorithm system dependent. Fig. \ref{FIG5}a shows the response of the digital cavity to sine functions calculated by the computer in the vicinity of 5 MHz for different values of $N_c$. The response at $N_c \leq 10^5$ shows that the signal amplitude is maximum for frequency at 5 MHz, which should be the case but the response at $N_c = 10^6$ shows that the maximum is shifted slightly by about 1 Hz towards lower frequency, which is an artifact. This artifact is related to the accuracy of the floating points in representing the actual numbers as well as the finite accuracy of the sine and cosine functions generated by the computer. Note, that this artifact is inherent in all the DSP systems as they use finite accuracy floating point system. Our tests show that using an Intel(R) Core(TM) i7-3770K CPU, gcc compiler (version 4.7) and 32-bit floating number system, we can achieve an accuracy of about of ppm (parts per million, Q-factor of about $ 10^5$) (Fig. \ref{FIG1}a), which is similar to the accuracy of the digital LIA found in the market.  As shown in Fig. \ref{FIG5}b, use of double precision (128-bit) number system improves the accuracy to few tens of ppb (parts per billion, Q-factor of about 10$^8$). In this case, we see the artifact due to inaccuracy of the number system and trigonometric function at $N_c = 10^9$ when the cavity is formed by unsing $n=20$. Here too, the artifact shows as the shift in the maximum response of the cavity to lower frequencies by about 2 mHz as shown in the inset Fig. \ref{FIG5}b. To our knowledge the Q-factor of $10^8$ that can be achieved with the generalized lock-in is far better than commercially available digital LIA. This makes the generalized LIA not only attractive for the precision analysis of high frequency signals but for any signals that modern ADCs are capable of digitizing.

\section{Conclusion}
In this work, we present a technique that allows us to generalize the concept of lock-in detection using the digital cavities. We call this technique the generalized lock-in. We have also shown that the generalized lock-in can be used to analyze high frequency signals (up to hundreds of MHz) with few tens of ppb resolution, and they can also be used to extract the signals from an extremely noisy data set. As the technique uses little computational footprint and is also parallelisable, we speculate that it is suitable for precision analysis of microwave signals upto 60-65 GHz that currently available digitizers can digitize. 

\textbf{Acknowledgments}

KK thanks Werner-Gren Foundation for the generous post-doctoral fellowship awarded to him.
Financial support from the Knut and Alice Wallenberg Foundation and Lund University Innovation System is greatfully acknowledged. 


%




\end{document}